\documentclass[11pt,a4paper]{article}
\usepackage[T1]{fontenc}
\usepackage[utf8]{inputenc}
\usepackage[english]{babel}
\usepackage[left=2.6cm,right=2.6cm,top=2.6cm,bottom=2.8cm]{geometry}
\usepackage{amsmath,amssymb,amsthm,mathtools}
\usepackage{mathptmx}
\usepackage{microtype}
\usepackage{graphicx}
\usepackage{placeins}
\usepackage{booktabs}
\usepackage{threeparttable}
\usepackage{array}
\usepackage{enumitem}
\usepackage{xcolor}
\usepackage[round,authoryear]{natbib}
\usepackage[colorlinks=true,linkcolor=blue!55!black,citecolor=blue!55!black,
            urlcolor=blue!55!black]{hyperref}

\numberwithin{equation}{section}
\newtheorem{theorem}{Theorem}
\newtheorem{corollary}{Corollary}[theorem]

\theoremstyle{definition}
\newtheorem{assumption}{Assumption}
\newtheorem{condition}{Condition}
\theoremstyle{remark}
\newtheorem{remark}{Remark}

\DeclareRobustCommand{\PN}{\ifmmode\text{\normalfont\textsc{pn}}\else{\normalfont\textsc{pn}}\fi}
\newcommand{\E}{\mathbb E}
\newcommand{\Pp}{\mathbb P}
\newcommand{\Q}{\mathbb Q}
\newcommand{\ind}{\mathbb I}

\DeclareMathOperator{\expit}{expit}

\setlist{nosep,leftmargin=*}
\graphicspath{{./}}

\title{\textbf{Retrospective Causal Attribution under Case--Control Sampling}}
\author{Zijian Sui\textsuperscript{1,2}\thanks{Email: \href{mailto:lemonade@mail.ustc.edu.cn}{\nolinkurl{lemonade@mail.ustc.edu.cn}}.},\quad
Hong Zhang\textsuperscript{1}\thanks{Email: \href{mailto:zhangh@ustc.edu.cn}{\nolinkurl{zhangh@ustc.edu.cn}}.},\quad
Jinfeng Xu\textsuperscript{3}\thanks{Email: \href{mailto:jinfenxu@cityu.edu.hk}{\nolinkurl{jinfenxu@cityu.edu.hk}}.}\quad and\quad
Min Zeng\textsuperscript{3}\thanks{Email: \href{mailto:zeng.min@my.cityu.edu.hk}{\nolinkurl{zeng.min@my.cityu.edu.hk}}.}\\[0.6em]
\parbox{0.95\textwidth}{\centering\small
\textsuperscript{1}Department of Statistics and Finance, School of Management,
University of Science and Technology of China, Anhui, China\\[0.3em]
\textsuperscript{2}Department of Systems Engineering, City University of Hong Kong,
Hong Kong\\[0.3em]
\textsuperscript{3}Department of Biostatistics, City University of Hong Kong,
Hong Kong, China}}
\date{}

\begin{document}
\maketitle

\begin{abstract}
	The probability of necessity (\PN) quantifies the probability that an exposed individual who experienced an outcome would not have experienced it in the absence of exposure. Case--control studies are an important resource for investigating etiologic questions, but their sampling design can introduce selection bias and complicate the statistical inference for \PN{}.
	In this paper, we develop a nonparametric framework for identification and efficient estimation of \PN{} under case--control sampling.
	With an externally supplied population outcome prevalence, we derive an exact identification formula that identifies \PN{} under standard causal assumptions and monotonicity and yields a valid lower bound without monotonicity.
	For rare outcomes, we derive a more tractable approximation that requires no external prevalence information and prove that its approximation error vanishes at the order of the population outcome prevalence.
	We further establish the semiparametric efficiency theory for the exact and approximate functionals, propose asymptotically efficient estimators, and construct confidence intervals for the corresponding targets.
	The proposed approach has potential applications in biomedical and epidemiological studies where causal attribution is investigated using retrospective data.
%The probability of necessity (\PN) quantifies the probability that an exposed individual who experienced an outcome would not have experienced it in the absence of exposure. Case--control studies are an important resource for investigating etiologic questions, but the sampling design can introduce selection bias for  identification and estimation of \PN{}.
% In this paper, we develop a nonparametric framework for identification and efficient estimation of \PN{} under case--control sampling.
%  First, we propose an general identification formula with an externally supplied population outcome prevalence. 
%  For uncommon outcomes, we derive a simpler approximation for \PN{} that requires no external prevalence information. The approximation can be quite accurate when the outcome is rare.
%  Then, we establish the semiparametric efficiency theory and propose novel
%  asymptotically efficient estimators of the two functionals, and further construct confidence intervals for \PN{} based on the proposed estimators.
%  An application to the Global Enteric Multicenter Study examines the attribution of moderate-to-severe diarrhea to Cryptosporidium and Rotavirus infections, with sensitivity analyses for the external prevalence.
\end{abstract}

%  For the exact and approximate functionals, we derive efficient influence functions in the retrospective model and construct cross-fitted estimators that allow flexible estimation of nuisance functions. 
%The estimators are asymptotically normal and efficient for their respective targets under suitable convergence-rate conditions. 
\section{Introduction}
\label{sec:introduction}

Causal attribution asks how likely an observed outcome was caused by an earlier exposure. 
A central measure is the probability of necessity (\PN), defined as the probability that the outcome would not have occurred without the exposure, given that the individual was exposed and experienced the outcome \citep{Pearl1999,TianPearl2000}. 
This measure is relevant to etiologic questions in epidemiology, such as whether diarrhea in an infected child would have been absent without the infection. 
Case--control studies provide a natural source of data for investigating such questions because they collect information from individuals with and without the outcome of interest. 
However, their outcome-dependent sampling design complicates causal attribution. 
Under sampling that depends only on the outcome, the case--control data preserve the distribution of exposure and covariates conditional on outcome status, but do not identify the outcome prevalence in the target population. 
Consequently, the population counterfactual risks needed to evaluate \PN{} cannot generally be recovered from the observed case--control distribution alone.

Classical work establishes identification and bounds for \PN{} using population distributions or causal risks \citep{Pearl1999,TianPearl2000,KurokiCai2011}. 
More recent work develops efficient inference under point identification \citep{tianwu2025} and inference for sharp covariate-assisted bounds without monotonicity in randomized experiments \citep{zhang2026}. 
A related literature studies causal and associational parameters under outcome-dependent sampling. 
Under a logistic disease model, \citet{prenticepyke1979} show that odds-ratio parameters and their asymptotic variances can be estimated by treating case--control data prospectively. 
For causal risk contrasts, \citet{gabriel2022} derive nonparametric bounds on the causal risk difference under several outcome-dependent sampling mechanisms with unmeasured confounding, while \citet{junlee2024} study causal relative and attributable risks under case--control and case--population sampling and derive sharp identified intervals under monotonicity assumptions. 
When the population outcome prevalence is known, \citet{vanderlaan2008} shows how prospective estimation procedures can be adapted to case--control designs. 
More recently, \citet{shingaki2026uai} identify probabilities of causation from case--control data without monotonicity using two proxy covariates or a proxy covariate together with an instrumental variable.

These results clarify several routes for causal inference under outcome-dependent sampling, but leave a practically important question for causal attribution among exposed cases. 
When reliable external information on the population outcome prevalence is available, how can it be used for efficient nonparametric inference on \PN{} under measured confounding? 
Conversely, when such prevalence information is unavailable, to what extent can \PN{} still be approximated from the case--control data alone when the outcome is rare? 
The second question is particularly relevant in applications where case--control studies are used precisely because the population outcome is uncommon, while reliable prevalence information may be difficult to obtain or transport across populations.

In this paper, we study these two routes to retrospective causal attribution. 
We first derive an exact identification functional $\theta_E$ from the observed case--control distribution when the population outcome prevalence $p_0$ is known. 
Under standard causal assumptions and monotonicity, $\theta_E$ identifies \PN{}; without monotonicity, it yields a valid lower bound. 
We then derive a prevalence-free functional $\theta_A$ for rare outcomes and show that, under uniform regularity conditions,
\[
\theta_A-\theta_E = O(p_0)
\]
as the population outcome prevalence $p_0$ tends to zero. 
We further establish the semiparametric efficiency theory for both functionals in a nonparametric model for case--control data and construct cross-fitted estimators based on their efficient influence functions. 
The resulting estimators accommodate flexible nuisance-function estimation and are asymptotically normal and efficient under suitable product-rate conditions. 
We evaluate their finite-sample performance in simulations and apply the proposed methods to the Global Enteric Multicenter Study \citep{kotloff2012protocol,kotloff2013}, where we assess the attribution of moderate-to-severe diarrhea to Cryptosporidium and Rotavirus infections and examine sensitivity to the external outcome prevalence.

The remainder of the paper is organized as follows. 
Section~2 develops identification of \PN{} under case--control sampling and introduces the rare-outcome approximation. 
Section~3 derives the efficient influence functions and presents the cross-fitted estimators and their asymptotic properties. 
Section~4 reports the simulation studies, and Section~5 presents the application to the Global Enteric Multicenter Study. 
Section~6 concludes with a discussion.
%The approximate functional is treated as a separate estimation target, allowing sampling uncertainty to be distinguished from approximation error.The influence functions are centered within cases and controls, accounting for outcome-stratified sampling.
\section{Causal attribution under retrospective sampling}
\label{sec:identification}

\subsection{Notations and basic setup}

We use the potential-outcome framework to define the causal attribution estimand. Suppose that \(n\) individuals are selected from a target population of interest under a retrospective sampling design. Let \(X\in\{0,1\}\) denote a binary exposure and let \(Y\in\{0,1\}\) denote a binary outcome. For each exposure level \(x\in\{0,1\}\), let \(Y_x\) denote the potential outcome that would have been observed had, possibly contrary to fact, the exposure been set to \(x\). Let \(Z\) denote a vector of pre-exposure confounders, and let \(S\in\{0,1\}\) indicate selection into the case--control sample, with \(S=1\) for sampled individuals. 
Let \(\Pp\) denote the joint distribution of \((Y_0,Y_1,Y,X,Z,S)\) in the target population. For the \(i\)th sampled individual (\(S_i=1\)), denote \(O_i=(Y_i,X_i,Z_i)\), \(i=1,\ldots,n\).
Let \(n_1=\#\{i:Y_i=1\}\) and \(n_0=\#\{i:Y_i=0\}\) denote the
observed numbers of cases and controls, respectively, so that \(n=n_1+n_0\).

The probability of necessity for causal attribution is defined as
\[
  \PN=\Pp(Y_0=0\mid X=1,Y=1).
\]
Among individuals who were exposed and experienced the outcome, \(\PN\) is the probability that the outcome would not have occurred had the exposure been absent.

Following the standard identification framework for probabilities of causation \citep{Pearl1999,TianPearl2000}, we impose the following assumptions for identification of \(\PN\).

\begin{assumption}[Consistency]
\label{ass:consistency}
\(Y=Y_X\) almost surely.
\end{assumption}

\begin{assumption}[Conditional exchangeability]
\label{ass:exchangeability}
\((Y_0,Y_1)\perp X\mid Z\).
\end{assumption}

\begin{assumption}[Positivity]
\label{ass:positivity}
\(\Pp(X=x\mid Z)>0\) almost surely for each exposure level \(x\) whose counterfactual mean is evaluated.
\end{assumption}

\begin{assumption}[Monotonicity]
\label{ass:monotonicity}
\(Y_1\ge Y_0\) almost surely.
\end{assumption}

Consistency links the observed outcome to the potential outcome under the observed exposure. Conditional exchangeability rules out unmeasured confounding after conditioning on \(Z\), while positivity ensures that the relevant exposure levels occur within the covariate strata of interest. Monotonicity excludes individuals for whom the exposure would prevent the outcome. Under Assumptions~\ref{ass:consistency}--\ref{ass:monotonicity}, if the data form a simple random sample from the target population, \(\PN\) can be identified and estimated using existing methods \citep{Pearl1999,TianPearl2000,KurokiCai2011}.

In this paper, we focus on identification of \(\PN\) under retrospective sampling. The following assumption characterizes the case--control sampling mechanism.
\begin{assumption}[Case--control sampling]
\label{ass:sampling}
Selection into the study depends only on the outcome; that is, \(S\perp(X,Z)\mid Y\), with \(\Pp(S=1\mid Y=y)>0\) for \(y\in\{0,1\}\).
\end{assumption}
For notational convenience, let $\Q$ denote the distribution of $O=(Y,X,Z)$ in the observed case--control sample, so that $\Q(\,\cdot\,)=\Pp(\,\cdot\mid S=1)$. Assumption~\ref{ass:sampling} implies $\Q(X,Z\mid Y=y)=\Pp(X,Z\mid Y=y)$ for $y\in\{0,1\}$. We write $\E_{\Pp}\{\cdot\}$ and $\E_{\Q}\{\cdot\}$ for expectations taken with respect to the target-population distribution $\Pp$ and the observed case--control distribution $\Q$, respectively.
%For notational convenience, define $Q(\,\cdot\,)=P(\,\cdot\mid S=1)$. Assumption~\ref{ass:sampling} implies $Q(X,Z\mid Y=y)=P(X,Z\mid Y=y)$ for $y\in\{0,1\}$. We write $\mathbb{E}_{P}\{\cdot\}$ and $\mathbb{E}_{Q}\{\cdot\}$ for expectations taken with respect to the target-population distribution and the observed case--control distribution, respectively.

In the next subsection, we develop a general theory for the nonparametric identification of \(\PN\) under case--control sampling. We then show that, when the outcome is rare, the identification formula for \(\PN\) admits a simple approximation that does not require external prevalence information.

\subsection{General identification for \PN{}}

Let \(p_0=\Pp(Y=1)\) denote the outcome prevalence in the target population,
let \(h_0=\Q(Y=1)\) denote the case fraction in the case--control sample, and
let \(\eta=\Pp(X=1\mid Y=1)=\Q(X=1\mid Y=1)\). Under i.i.d. sampling from
\(\Q\), \(\widehat h_0=n_1/n\) estimates \(h_0\). Under fixed-stratum
sampling, \(n_1\) and \(n_0\) are set by design, and we take
\(h_0=n_1/n\). Denote \(\mu_0=\Pp(Y_0=1)\) and define the \emph{Exact}
functional by \(\theta_E:=(p_0-\mu_0)/(p_0\eta)\).
The following theorem presents an nonparametric identification result for $\PN$ in the case-control setting.
%The external \(p_0\) is treated as known information. 

\begin{theorem}[Identification of \PN{} in case--control studies]
\label{thm:exact-identification}
Suppose Assumptions~\ref{ass:consistency}, \ref{ass:exchangeability}, \ref{ass:positivity}, and \ref{ass:sampling} hold.
\begin{enumerate}[label=(\alph*)]
  \item Under Monotonicity (Assumption~\ref{ass:monotonicity}), \(\PN=\theta_E\).
  \item Without Monotonicity, \(\PN\geq\max\{0,\theta_E\}\).
\end{enumerate}
Moreover, for \(x\in\{0,1\}\), the counterfactual risk is identified by
\begin{equation}
\Pp(Y_x=1)
=\E_{\Pp}\left[
\frac{r(Z,p_0)\pi(x\mid1,Z)}
{\pi(x\mid0,Z)+r(Z,p_0)\{\pi(x\mid1,Z)-\pi(x\mid0,Z)\}}
\right].
\label{eq:counterfactual-risk}
\end{equation}
where \(\pi(x\mid y,z):=\Q(X=x\mid Y=y,Z=z)\), and
\[
r(z,p):=
\frac{p(1-h_0)\Q(Y=1\mid Z=z)}
{p(1-h_0)\Q(Y=1\mid Z=z)+h_0(1-p)\Q(Y=0\mid Z=z)}.
\]
In particular, \(\mu_0=\Pp(Y_0=1)\) is obtained from Eq.~\eqref{eq:counterfactual-risk} by setting \(x=0\).
\end{theorem}

Theorem~\ref{thm:exact-identification} gives the general identification result. It provides an exact, nonparametric identification formula for \(\PN\). A key challenge in applying Theorem \ref{thm:exact-identification} is the estimation of the counterfactual probability \(\Pp(Y_x=1)\) from observed data, since the retrospective sampling design distorts the population distribution. To correct for this, we use inverse probability of sampling weights (IPSW), defined for each sampled individual \(i\) by
\begin{equation}
w(Y_i)=\mathbb I(Y_i=1)\frac{p_0}{h_0}
+\mathbb I(Y_i=0)\frac{1-p_0}{1-h_0}.
\label{eq:ipsw}
\end{equation}
Here \(\mathbb I(\cdot)\) denotes the indicator function. For any integrable function \(g(Y,X,Z)\), Assumption~\ref{ass:sampling} gives \(\E_{\Pp}\{g(Y,X,Z)\}=\E_{\Q}\{w(Y)g(Y,X,Z)\}\).  Therefore, \(\theta_E\) is identified from the case--control data with a known outcome prevalence \(p_0\).

\begin{remark}
In addition to the identification strategy in Theorem~\ref{thm:exact-identification}, several standard approaches may be used to estimate the counterfactual mean \(\Pp(Y_x=1)\) through suitable weighting or outcome modeling. Common examples include regression imputation, inverse probability weighting, and augmented inverse probability weighting \citep{horvitz1952generalization,rosenbaum1983central,bang2005doubly}.
\end{remark}

\begin{remark}
	Under the case-control sampling design, the selection indicator $S$ is influenced by $Y$. To address the selection biases of our $\PN$ estimators, the disease prevalence \(p_0\) should be correctly specified or estimated based on external knowledge.
	When \(f\) is estimated with high precision, its uncertainty is expected to have a negligible impact on the $\PN$ estimators. The uncertainty of the estimated \(p_0\) can also be incorporated straightforwardly by the delta method or bootstrapping.
\end{remark}

\subsection{A rare-outcome approximation for \PN{} without external prevalence}
\label{sec:rare-approximation}

The Exact identification functional $\theta_E$ has two practical limitations. First, its form is relatively complicated, making computation and estimation somewhat cumbersome in practice. Second, its evaluation requires external information $p_0$. In this subsection, we show that, when the outcome is rare ($p_0 \approx 0$), $\theta_E$ admits a substantially simpler approximation and is therefore more convenient to estimate without external prevalence information.

%\begin{corollary}[Rare-outcome approximation for \PN{}]
%\label{cor:rare-approximation}
%Define the \emph{Approx} functional
%\begin{equation}
%\theta_A:=\frac{1}{\eta}\left[1-
%\E_{\Q}\left\{
%\frac{1-h_0}{h_0}
%\frac{\Q(Y=1\mid Z)}{\Q(Y=0\mid Z)}
%\frac{\pi(0\mid1,Z)}{\pi(0\mid0,Z)}
%\ \middle|\ Y=0
%\right\}\right].
%\label{eq:theta-approx}
%\end{equation}
%Under Assumptions~\ref{ass:consistency}, \ref{ass:exchangeability}, \ref{ass:positivity}, and \ref{ass:sampling}, suppose that $p_0 \approx 0$, we have
%\begin{enumerate}
%	\item[\textbf{(a)}] \textbf{Under Monotonicity (Assumption \ref{ass:monotonicity}):} $\PN \approx \theta_A$.
%	\item[\textbf{(b)}] \textbf{Without Monotonicity (Assumption \ref{ass:monotonicity}):}  $\max\{0,\theta_A\}$ is a rare-outcome approximation to the valid lower bound
%	$\max\{0,\theta_E\}$ for $\PN$.
%\end{enumerate}
%\end{corollary}
\begin{corollary}[Rare-outcome approximation for \PN{}]
	\label{cor:rare-approximation}
	Define the \emph{Approx} functional
	\begin{equation}
		\theta_A:=\frac{1}{\eta}\left[1-
		\E_{\Q}\left\{
		\frac{1-h_0}{h_0}
		\frac{\Q(Y=1\mid Z)}{\Q(Y=0\mid Z)}
		\frac{\pi(0\mid1,Z)}{\pi(0\mid0,Z)}
		\ \middle|\ Y=0
		\right\}\right].
		\label{eq:theta-approx}
	\end{equation}
	Consider a sequence of target-population distributions $\Pp_{p_0}$ with $p_0=\Pp_{p_0}(Y=1)\downarrow0$. Let $\theta_{E,p_0}$, $\theta_{A,p_0}$, and $\PN_{p_0}$ denote the Exact functional, the Approx functional, and the probability of necessity under $\Pp_{p_0}$, respectively. Suppose that each \(\Pp_{p_0}\) satisfies
	Assumptions~\ref{ass:consistency}, \ref{ass:exchangeability},
	\ref{ass:positivity}, and \ref{ass:sampling}, and that the regularity
	conditions stated in the Supplementary Material hold uniformly along the sequence. Then
	\begin{enumerate}
		\item[\textbf{(a)}] \textbf{Under Monotonicity:} 
$$
\theta_{A,p_0}-\PN_{p_0}\longrightarrow0.
$$
		\item[\textbf{(b)}] \textbf{Without Monotonicity:}
		$\max\{0,\theta_A\}$ is a rare-outcome approximation to the valid lower bound
			$\max\{0,\theta_E\}$ for $\PN$:
			$$
		\left|\max\{0,\theta_{A,p_0}\}-\max\{0,\theta_{E,p_0}\}\right|
		\longrightarrow0.
		$$
	\end{enumerate}
\end{corollary}

Under monotonicity, \(\theta_A\) serves as a prevalence-free rare-outcome approximation to \(\PN=\theta_E\). Without monotonicity, \(\theta_A\) approximates the untruncated smooth lower-bound of $\PN$ due to \(\theta_A \approx \theta_E\). The following theorem gives the approximation rate.

\begin{theorem}[Rare-outcome approximation rate]
\label{thm:rare-expansion}
Consider a sequence of target-population distributions \(\Pp_{p_0}\), indexed by
\(p_0=\Pp_{p_0}(Y=1)\downarrow0\). Let \(\theta_{E,p_0}\) and
\(\theta_{A,p_0}\) denote the Exact and Approx functionals under \(\Pp_{p_0}\),
respectively. Suppose that each \(\Pp_{p_0}\) satisfies
Assumptions~\ref{ass:consistency}, \ref{ass:exchangeability},
\ref{ass:positivity}, and \ref{ass:sampling}, and that the regularity
conditions stated in the Supplementary Material hold uniformly along the sequence. Then
\begin{equation}
\theta_{A,p_0}-\theta_{E,p_0}=O(p_0).
\label{eq:rare-remainder}
\end{equation}
\end{theorem}

Under these regularity conditions, the approximation error is $O(p_0)$ as $p_0\downarrow0$.

\begin{remark}
\label{rem:approximation-gap}
The two functionals admit the equivalent representations
$$
\begin{aligned}
\theta_E
&=\frac{1}{\eta}\left[1-\E_{\Q}\left\{\frac{\pi(0\mid1,Z)}{\Pp(X=0\mid Z)}\,\middle|\,Y=1\right\}\right],\\
\theta_A
&=\frac{1}{\eta}\left[1-\E_{\Q}\left\{\frac{\pi(0\mid1,Z)}{\pi(0\mid0,Z)}\,\middle|\,Y=1\right\}\right].
\end{aligned}
$$
Consequently,
$$
\theta_A-\theta_E
=\frac{1}{\eta}\E_{\Q}\left[\Pp(Y=1\mid X=0,Z)\left\{1-\frac{\pi(0\mid1,Z)}{\pi(0\mid0,Z)}\right\}\,\middle|\,Y=1\right].
$$
Under conditional exchangeability and monotonicity, $\Pp(Y=1\mid X=1,Z)\geq \Pp(Y=1\mid X=0,Z)$, which implies $\pi(0\mid1,Z)\leq\pi(0\mid0,Z)$ by Bayes' rule and Assumption~\ref{ass:sampling}. Hence $\theta_A\geq\theta_E=\PN$. Moreover, if $\Pp(Y=1\mid X=0,Z)\leq\varepsilon$ for some $0<\varepsilon<1$, then under monotonicity $0\leq\theta_A-\theta_E\leq\varepsilon\theta_A$. Therefore, uniformly small unexposed conditional risks $\Pp(Y=1\mid X=0,Z)$ control the approximation error.
%a small marginal prevalence alone does not ensure this conditional-risk bound. 
%Without monotonicity, the gap need not have a fixed sign.
%over the covariate distribution among cases 
\end{remark}

\section{Efficient estimation in the retrospective model}
\label{sec:estimation}

In this section, we present the efficient influence functions (EIFs) and
efficiency bounds for \(\theta_E\) and \(\theta_A\), and then use the EIFs to
construct cross-fitted estimators of the two functionals.

\subsection{Efficient influence functions}
\label{sec:efficient-influence-functions}

An efficient estimator attains the semiparametric efficiency bound, which represents the
smallest asymptotic variance among regular asymptotically linear estimators in the specified statistical model.
In this paper, we denote the specified model by \(\mathcal M\), so that \(\Q \in \mathcal M\). Within \(\mathcal M\), the conditional distributions of \((X,Z)\), $\Q(X,Z \mid Y = y),\; y =0,1$ are unrestricted. 
%The case fraction \(h_0 = Q(Y=1)\) is determined by sampling design, which is ancillary for both target functionals. 

We first derive the EIF and efficiency bound for
each target functional. Let $\nu(Z)=\{m(Z),\pi(0\mid1,Z),\pi(0\mid0,Z)\}$
denote the nuisance parameter vector for \(\theta_E\) and \(\theta_A\), where \(m(z)=\Q(Y=1\mid Z=z)\).
We use the notation \(\pi(x\mid y,z)\) and \(r(z,p_0)\) defined in
Section~\ref{sec:identification}. For fixed \(p_0\) and \(h_0\), define 
\[
d_\nu(z)
=\pi(0\mid0,z)
+r(z,p_0)\{\pi(0\mid1,z)-\pi(0\mid0,z)\},
\qquad
q_\nu(z)
=\frac{r(z,p_0)\pi(0\mid1,z)}{d_\nu(z)}.
\]
The functions \(d_\nu\) and \(q_\nu\) are algebraic transforms of the nuisance parameters.
Define
\begin{equation}
 H_E(O;\nu)
 =\frac{\ind(X=0)}{d_\nu(Z)}\{Y-q_\nu(Z)\}+q_\nu(Z),
 \qquad c_y=\E_{\Q}\{H_E(O;\nu)\mid Y=y\}.
 \label{eq:exact-signal}
\end{equation}

%The external prevalence \(p_0\) is treated as fixed in the EIF analysis. The case fraction \(h_0\) is ancillary: it is fixed by design under fixed-stratum sampling and can be replaced by its empirical counterpart under the i.i.d. mixture representation. 
\begin{theorem}[Efficient influence functions]
\label{thm:eifs}
Suppose Assumptions~\ref{ass:consistency}, \ref{ass:exchangeability},
\ref{ass:positivity}, and \ref{ass:sampling} hold and the regularity
conditions in Supplementary Section S1.3 are satisfied. Then the efficient
influence functions for \(\theta_E\) and \(\theta_A\) are
\begin{align}
 D_E(O;\nu)
 &=-\frac{w(Y)}{p_0\eta}
 \left[H_E(O;\nu)-c_Y+\theta_E\ind(Y=1)(X-\eta)\right],
 \label{eq:eif-exact}\\
 D_A(O;\nu)
 &=\frac{1}{h_0\eta}\left[
 \ind(Y=1)\left\{1-\theta_A X-
 \frac{1-X}{\pi(0\mid0,Z)}\right\}
 \right.\nonumber\\
 &\hspace{24mm}\left.
 +\ind(Y=0)\frac{m(Z)}{1-m(Z)}
 \frac{\pi(0\mid1,Z)}{\pi^2(0\mid0,Z)}
 \{1-X-\pi(0\mid0,Z)\}
 \right].
 \label{eq:eif-approx}
\end{align}
For \(j\in\{E,A\}\), \(\E_{\Q}\{D_j(O;\nu)\mid Y=y\}=0\) for \(y=0,1\).
\end{theorem}

%\begin{remark}
%\label{rem:eif-without-monotonicity}
%Theorem~\ref{thm:eifs} does not require monotonicity. Under Assumption~\ref{ass:monotonicity}, $D_E(O;\nu)$ is the EIF for $\PN$, whereas $D_A(O;\nu)$ is the EIF for its rare-outcome approximation $\theta_A$. Without monotonicity, $\PN$ is generally only partially identified, and consistency gives $\theta_E=\PN-\frac{\Pp(Y_0=1,Y_1=0,X=1)}{\Pp(X=1,Y=1)}$. 
%%In this case, $D_E(O;\nu)$ is the EIF for the smooth, untruncated lower-bound functional $\theta_E$, while $D_A(O;\nu)$ is the EIF for $\theta_A$, which need not itself bound $\PN$. The bound $\max\{0,\theta_E\}$ is not generally sharp and is generally nonregular at $\theta_E=0$. 
%In randomized experiments, \citet{zhang2026} prove the sharpness of covariate-assisted bounds and develop efficient inference under a margin condition for these nonsmooth functionals. A natural extension is to express sharp lower and upper bounds using the retrospective conditional distributions and external prevalence, and to develop inference that addresses the nonsmoothness of the covariate-specific risk comparisons.
%\end{remark}

For an i.i.d. sample of size \(n\) drawn from \(\Q\), the efficiency
bound for \(\sqrt n(\widehat\theta_j-\theta_j)\) is
\(\E_{\Q}\{D_j(O;\nu)^2\}\), \(j\in\{E,A\}\). 
%Under fixed stratum sizes,
%\(h_0=n_1/n\) is the design case fraction. Let \(\omega_1=h_0\),
%\(\omega_0=1-h_0\), and
%\(\phi_{jy}(x,z)=\omega_yD_j(y,x,z;\nu)\). The first-order variance is
%\begin{equation}
% \frac{V_{j1}}{n_1}+\frac{V_{j0}}{n_0},
% \qquad
% V_{jy}=\operatorname{Var}_{Q}\{\phi_{jy}(X,Z)\mid Y=y\}.
% \label{eq:two-sample-variance}
%\end{equation}

\begin{remark}
	\label{rem:case-control-sampling}
	The case fraction $h_0=\Q(Y=1)$ is induced by the sampling design and is ancillary to both target functionals. Under Assumption~\ref{ass:sampling}, researchers may obtain the case--control sample under either of two sampling schemes. First, cases and controls may be independently sampled according to the design-specified probabilities $\Pp(S=1 \mid Y=1)$ and $\Pp(S=1 \mid Y=0)$. In this setting, an empirical estimator of $h_0$ is $\widehat h_0=n_1/n$. By the central limit theorem,
	$\widehat h_0-h_0=O_p(n^{-1/2})$. Second, the numbers of sampled cases and controls, $n_1$ and $n_0$, may be fixed by design, with observations sampled independently within each outcome stratum $Y = y$. In this setting, $h_0=n_1/n$ is a known design constant. 
	
%	Under Assumption~\ref{ass:sampling}, researchers may obtain the case--control sample under either of two sampling schemes. First, cases and controls may be randomly sampled according to the design-specified probabilities $P(S=1 \mid Y=1)$ and $P(S=1 \mid Y=1)$. In this setting, an empirical estimator of $h_0$ is $\widehat h_0=n_1/n$, which satisfies $\widehat h_0-h_0=O_p(n^{-1/2})$. The EIFs in Theorem~\ref{thm:eifs} are orthogonal to the score for $h_0$. Consequently, estimating $h_0$ does not contribute an additional first-order variance component. Second, the numbers of sampled cases and controls, $n_1$ and $n_0$, may be fixed by design. In this setting, $h_0=n_1/n$ is a known design constant, and hence $\widehat h_0=h_0=n_1/n$. Accordingly, in the estimation procedures below, we use $n_1/n$ in place of $h_0$.
\end{remark}

%Under fixed stratum sizes,
%\(h_0=n_1/n\) is the design case fraction. Let \(\omega_1=h_0\),
%\(\omega_0=1-h_0\), and
%\(\phi_{jy}(x,z)=\omega_yD_j(y,x,z;\nu)\). The first-order variance is
%\begin{equation}
% \frac{V_{j1}}{n_1}+\frac{V_{j0}}{n_0},
% \qquad
% V_{jy}=\operatorname{Var}_{Q}\{\phi_{jy}(X,Z)\mid Y=y\}.
% \label{eq:two-sample-variance}
%\end{equation}
%We use this variance in the simulation study.

%The EIF for \(\theta_E\) also follows from the prospective augmented EIF for
%\(\PN\) through the known-prevalence case--control transformation. Applying
%that transformation gives Eq.~\eqref{eq:eif-exact}. Supplementary Section S1.8
%maps our notation to the prospective result of \citet{tianwu2025} and the
%known-prevalence framework of \citet{vanderlaan2008}. This representation
%motivates the cross-fitted construction below. We additionally derive the
%prevalence-free functional \(\theta_A\) and state its approximation error
%explicitly.

\subsection{Cross-fitted estimation}

In this subsection, we propose efficient estimators for $\theta_E$ and $\theta_A$. Denote $\Q_nf=\frac{1}{n}\sum_{i=1}^n f(O_i)$. The computational algorithm is given below.

\paragraph{Step 1: Sample statistics and outer folds.}
Let
\[
\widehat h_0=\frac{n_1}{n},
\qquad
\widehat\eta=\frac{1}{n_1}\sum_{Y_i=1}X_i.
\]
%Under fixed-stratum sampling, \(\widehat h_0=h_0\) by design; under the i.i.d.
%mixture representation, \(\widehat h_0\) is the empirical estimator of
%\(h_0=Q(Y=1)\).
We partition the index set \(\{1,\ldots,n\}\) into a fixed number \(K\) of
mutually disjoint validation folds \(I_1,\ldots,I_K\), stratified by outcome,
such that \(\bigcup_{k=1}^K I_k=\{1,\ldots,n\}\). For fold \(k\), \(I_k\) is
the validation index set and
\(I_k^c=\{1,\ldots,n\}\setminus I_k\) is the corresponding outer-training
index set. The nuisance functions are fitted using
\(\{O_i:i\in I_k^c\}\) and evaluated for \(i\in I_k\).

\paragraph{Step 2: Fit the nuisance functions.}
For outer fold \(k\), we fit \(m(z)=\Q(Y=1\mid Z=z)\) using all outer-training
observations, \(\pi(0\mid1,z)=\Q(X=0\mid Y=1,Z=z)\) using the outer-training
cases, and \(\pi(0\mid0,z)=\Q(X=0\mid Y=0,Z=z)\) using the outer-training
controls. Denote the resulting fold-specific nuisance parameter vector by
\[
\widehat\nu^{(-k)}(Z)
=\{\widehat m^{(-k)}(Z),
\widehat\pi^{(-k)}(0\mid1,Z),
\widehat\pi^{(-k)}(0\mid0,Z)\}.
\]
An estimated nuisance function evaluated
at \(Z_i,\; i\in I_k\) denotes the fit trained on \(I_k^c\). For notational convenience, we suppress the superscript \((-k)\).
For a validation observation \(i\in I_k\), compute
\begin{align*}
\widehat r_i
&=\frac{p_0(1-\widehat h_0)\widehat m(Z_i)}
{p_0(1-\widehat h_0)\widehat m(Z_i)
+\widehat h_0(1-p_0)\{1-\widehat m(Z_i)\}},\\
\widehat d_i
&=d_{\widehat\nu^{(-k)}}(Z_i)\\
&=\widehat\pi(0\mid0,Z_i)
+\widehat r_i\{\widehat\pi(0\mid1,Z_i)
-\widehat\pi(0\mid0,Z_i)\},\\
\widehat q_i
&=q_{\widehat\nu^{(-k)}}(Z_i)
=\frac{\widehat r_i\widehat\pi(0\mid1,Z_i)}{\widehat d_i}.
\end{align*}
%Thus \(\widehat d_i\) and \(\widehat q_i\) are not obtained from separately
%trained learners; they are algebraic plug-in transforms of
%the functions collected in \(\widehat\nu^{(-k)}(\cdot)\). Any probability
%truncation is prespecified as a
%numerical safeguard and is applied to nuisance predictions, not to either
%final functional estimate.

\paragraph{Step 3: Compute \(\widehat\theta_E\).}
Let
\[
\widehat H_{E,i}
=\frac{\ind(X_i=0)}{\widehat d_i}(Y_i-\widehat q_i)+\widehat q_i, \quad \widehat\mu_0^{\mathrm{aug}}=\frac{p_0}{n_1}\sum_{Y_i=1}\widehat H_{E,i}
+\frac{1-p_0}{n_0}\sum_{Y_i=0}\widehat H_{E,i}.
\]
The closed-form calibration root is obtained from
\begin{align}
 \widehat\theta_E
 =\frac{p_0-\widehat\mu_0^{\mathrm{aug}}}
 {p_0\widehat\eta}.
\end{align}
%\begin{align}
%	\widehat\mu_0^{\mathrm{aug}}
%	&=\frac{p_0}{n_1}\sum_{Y_i=1}\widehat H_{E,i}
%	+\frac{1-p_0}{n_0}\sum_{Y_i=0}\widehat H_{E,i},
%	\label{eq:mu-aug-estimator}\\
%	\widehat\theta_E
%	&=\frac{p_0-\widehat\mu_0^{\mathrm{aug}}}
%	{p_0\widehat\eta}.
%	\label{eq:exact-estimator}
%\end{align}
%The estimates \(\widehat c_y\) enter only the EIF-based variance estimate and
%not the point estimate.

\paragraph{Step 4: Compute \(\widehat\theta_A\).}
The corresponding estimating-equation root has the closed form
\begin{align}
\widehat\theta_A
&=\frac{1}{\Q_n\{\ind(Y=1)X\}}
\Q_n\Bigg[
\ind(Y=1)\left\{1-
\frac{1-X}{\widehat\pi(0\mid0,Z)}\right\}
\nonumber\\
&\hspace{26mm}
+\ind(Y=0)
\frac{\widehat m(Z)}{1-\widehat m(Z)}
\frac{\widehat\pi(0\mid1,Z)}
{\widehat\pi^2(0\mid0,Z)}
\left\{1-X-\widehat\pi(0\mid0,Z)\right\}
\Bigg].
\label{eq:approx-estimator}
\end{align}
All estimated functions in Eq.~\eqref{eq:approx-estimator} are the
out-of-fold predictions from Step 2. 

%Because
%\(Q_n\{\ind(Y=1)X\}=\widehat h_0\widehat\eta\), this expression is
%algebraically identical to an EIF estimating equation. Specifically, for
%\(i\in I_k\), let \(\widehat D_A(O_i;\theta)\) be the cross-fitted estimating
%function obtained from Eq.~\eqref{eq:eif-approx} by replacing \(\theta_A\)
%with \(\theta\) and all nuisance and scalar quantities by their out-of-fold
%or empirical estimates. Then Eq.~\eqref{eq:approx-estimator} is the solution
%to
%\[
% \frac{1}{n}\sum_{i=1}^n \widehat D_A(O_i;\theta)=0.
%\]

%\paragraph{Step 5: Estimate EIF variances and form Wald intervals.}
%For \(i\in I_k\) and \(j\in\{E,A\}\), let \(\widehat D_{j,i}\) denote
%the cross-fitted plug-in version of \(D_j(O_i;\nu)\), obtained by replacing
%\(\theta_j\) with \(\widehat\theta_j\), \(\nu\) with
%\(\widehat\nu^{(-k)}\), and \(h_0\), \(\eta\), and \(c_y\), as applicable,
%by their corresponding empirical estimates; \(p_0\) remains fixed. Thus
%\(\widehat D_{A,i}=\widehat D_A(O_i;\widehat\theta_A)\). In
%\(\widehat D_{E,i}\), \(c_y\) is estimated within the corresponding
%outer-training outcome stratum and is used only for variance estimation. Under fixed
%\(n_1,n_0\), we use Eq.~\eqref{eq:two-sample-variance}; the Wald intervals use
%the plug-in sample variances of the cross-fitted EIF contributions. Both roots
%are left unprojected. Projection to \([0,1]\), or replacement of \(\theta_E\)
%by \(\max(0,\theta_E)\), changes the regularity analysis at the boundary.

\subsection{Sampling asymptotics and variance estimation}

Let $\|\cdot\|_2$ denote the $L_2$-norm with respect to $\Q$. Let
$\widehat\nu$ denote the nuisance-function estimates obtained in Step~2, and write
$
\widehat d=d_{\widehat\nu},\,
\widehat q=q_{\widehat\nu}.
$

\begin{condition}[Product-rate conditions]
\label{cond:product-rates}
Uniformly across folds, assume
\begin{equation}
 \|\widehat d-d_\nu\|_2
 \|\widehat q-q_\nu\|_2=o_p(n^{-1/2}).
 \label{eq:rate-exact}
\end{equation}
For $\hat{\theta}_A$, assume
\begin{align}
&\|\widehat\pi(0\mid0,\cdot)-\pi(0\mid0,\cdot)\|_2
\Big\{
\|\widehat\pi(0\mid0,\cdot)-\pi(0\mid0,\cdot)\|_2
+\|\widehat m-m\|_2 \nonumber\\
&\hspace{37mm}
+\|\widehat\pi(0\mid1,\cdot)-\pi(0\mid1,\cdot)\|_2
 \Big\}=o_p(n^{-1/2}).
 \label{eq:rate-approx}
\end{align}
\end{condition}

Condition~\ref{cond:product-rates} allows the component nuisance errors to
converge at unequal rates. A simple sufficient scenario is that both
\(\widehat d\) and \(\widehat q\) converge faster than \(n^{-1/4}\) for
\(\theta_E\), and that the three fitted retrospective regressions converge
faster than \(n^{-1/4}\) for \(\theta_A\). 
%These are sufficient examples rather
%than imposed componentwise rates; cross-fitting itself does not guarantee them.

\begin{theorem}[Asymptotic normality and efficiency]
	\label{thm:asymptotic}
	Suppose Assumptions~\ref{ass:consistency}, \ref{ass:exchangeability},
	\ref{ass:positivity}, and \ref{ass:sampling},
	Condition~\ref{cond:product-rates}, and the regularity conditions in
	Supplementary Sections S1.3 and S1.6 hold. 
	\begin{itemize}
		\item for the Exact estimator,
		\begin{equation*}
			\sqrt{n}\{\widehat\theta_E-\theta_E\}
			\xrightarrow{d}
			\mathcal N(0,\sigma_E^2),
			\qquad
			\sigma_E^2=\E_{\Q}\{D_E(O;\nu)^2\};
		\end{equation*}
		\item for the Approx estimator,
		\begin{equation*}
			\sqrt{n}\{\widehat\theta_A-\theta_A\}
			\xrightarrow{d}
			\mathcal N(0,\sigma_A^2),
			\qquad
			\sigma_A^2=\E_{\Q}\{D_A(O;\nu)^2\}.
		\end{equation*}
	\end{itemize}
	Here $\xrightarrow{d}$ denotes convergence in distribution. For
	$j\in\{E,A\}$, $\sigma_j^2$ is the semiparametric efficiency bound for
	$\theta_j$.
\end{theorem}

For $i\in I_k$ and $j\in\{E,A\}$, let $\widehat D_{j,i}$ denote a consistent estimator of $D_j(O_i;\nu)$ obtained by replacing
$\theta_j$ with $\widehat\theta_j$, $\nu$ with
$\widehat\nu^{(-k)}$, and $\eta$ with $\widehat\eta$. A consistent estimator of $\sigma_j^2$ is
\begin{equation}
	\widehat\sigma_j^2
	=n^{-1}\sum_{i=1}^n
	\left\{
	\widehat D_{j,i}
	-n^{-1}\sum_{\ell=1}^n\widehat D_{j,\ell}
	\right\}^2,
	\qquad j\in\{E,A\}.
	\label{eq:estimated-eif-variance}
\end{equation}
%Under the conditions of Theorem~\ref{thm:asymptotic},
%$\widehat\sigma_j^2\xrightarrow{p}\sigma_j^2$ under either sampling scheme.
%Accordingly, the estimated standard error is
%\begin{equation*}
%	\widehat{\mathrm{se}}_j
%	=\sqrt{\frac{\widehat\sigma_j^2}{n}},
%\end{equation*}
%and 
A two-sided $100(1-\alpha)\%$ Wald confidence interval for $\theta_j$ is
$\widehat\theta_j
	\mathbin{\pm}
	z_{1-\alpha/2}\sqrt{\frac{\widehat\sigma_j^2}{n}},$
where $z_{1-\alpha/2}$ is the $1-\alpha/2$ quantile of the standard normal
distribution.
\begin{remark}
	\label{rem:estimand-choice}
	Researchers should choose the appropriate functional according to the practical setting. When reliable external information on $p_0$ is available, researchers can use the Exact estimator $\widehat\theta_E$ for inference on $\theta_E$.
	In the absence of such information,, the Approx estimator $\widehat\theta_A$ can provide a rare-outcome approximation to $\widehat\theta_E$ and help assess the influence of prevalence specification by comparison with $\widehat\theta_E$ evaluated at plausible working values of $p_0$. Close agreement between $\widehat\theta_E$ and $\widehat\theta_A$ suggests that the prevalence correction has a limited effect on estimation.
\end{remark}

\begin{remark}
	\label{rem:estimand-A}
	An important question is how to assess the reliability of $\widehat\theta_A$. Its deviation from the Exact target $\theta_E$ can be decomposed as $\widehat\theta_A-\theta_E=(\widehat\theta_A-\theta_A)+(\theta_A-\theta_E)$, where $\widehat\theta_A-\theta_A$ and $\theta_A-\theta_E$ represent estimation error and approximation bias, respectively. Under the conditions of Theorem~\ref{thm:asymptotic}, the estimation error $\widehat\theta_A-\theta_A$ converges to zero in probability as the sample size increases, whereas the approximation bias $\theta_A-\theta_E$ is a systematic difference that does not change with sample size for a fixed population. The approximation bias vanishes either as $p_0\to0$, or as $\Pp(Y=1\mid X=0,Z)$ tends uniformly to zero under monotonicity assumption. Furthermore, our proposed estimators $\widehat\theta_A$ and
	$\widehat\theta_E$ are jointly asymptotically normal:
	$$
	\sqrt n
	\begin{pmatrix}
		\widehat\theta_A-\theta_A\\
		\widehat\theta_E-\theta_E
	\end{pmatrix}
	\xrightarrow{d}
	\mathcal N_2\left\{
	\boldsymbol 0,
	\begin{pmatrix}
		\sigma_A^2 & \sigma_{AE}\\
		\sigma_{AE} & \sigma_E^2
	\end{pmatrix}
	\right\},
	$$
	where $\sigma_{AE}=\E_Q\{D_A(O;\nu)D_E(O;\nu)\}$. Thus, $\theta_A-\theta_E$ can be estimated by $\widehat\theta_A-\widehat\theta_E$, with estimated standard error $\{(\widehat\sigma_A^2+\widehat\sigma_E^2-2\widehat\sigma_{AE})/n\}^{1/2}$, where $\widehat\sigma_{AE}$ is a consistent estimator of $\sigma_{AE}$.
	
%	the empirical covariance of the paired cross-fitted EIF contributions.  Provided $\sigma_A^2+\sigma_E^2-2\sigma_{AE}>0$, the corresponding Wald confidence interval can be used to assess the magnitude of the approximation bias.
\end{remark}
\section{Simulation study}
\label{sec:simulation}

We conduct simulations to evaluate the finite-sample performance of the Exact and Approx estimators under case--control sampling. The data-generating mechanism includes nonlinear interactions and measured confounding.

The data generation comprises two steps: data pool creation and case-control sample selection. We start by creating a data pool representing the target population $\mathbb{P}$ with a large size of $N=\left\lceil 200000/p_0\right\rceil$. We independently generate a discrete covariate $Z_1\sim\operatorname{Bernoulli}(0.5)$ and a continuous covariate $Z_2\sim N(5,1)$, and define $U=Z_2-5$. The exposure indicator is sampled from a Bernoulli distribution with success probability $\Pp(X=1\mid Z)=\expit{1+0.8Z_1-0.2Z_2+0.2Z_1U}$.

\begingroup
\emergencystretch=1em
To represent heterogeneity in causal responses, we introduce a principal-stratum variable
$G\in\{\mathrm{never},\mathrm{causal},\mathrm{always},\mathrm{defier}\}$.
The never, causal, and always strata correspond to
$(Y(0),Y(1))=(0,0),(0,1)$, and $(1,1)$, respectively,
whereas the defier stratum corresponds to $(1,0)$.
We set $\mathbb{P}(G=\mathrm{defier})=0$ to ensure monotonicity.

For $g\in\mathcal{G}=\{\mathrm{never},\mathrm{causal},\mathrm{always}\}$,
let $\eta_g(Z)=\alpha+\gamma_{g0}+\gamma_{g1}Z_1+\gamma_{g2}Z_2+\gamma_{g3}Z_1U$.
The conditional response-type probabilities are
\[
\mathbb{P}(G=g\mid Z)
=\frac{\mathbb{I}\{g=\mathrm{never}\}+\exp\{\eta_g(Z)\}}
{1+\sum_{h\in\mathcal{G}}\exp\{\eta_h(Z)\}},
\qquad g\in\mathcal{G}.
\]
The coefficient vectors $(\gamma_{g0},\ldots,\gamma_{g3})$ are
$(-2.0,0.1,0,0.1)$ for never, $(-3.0,0.2,0.1,0.1)$ for causal,
and $(-1.0,0,-0.1,-0.1)$ for always.
The common intercept $\alpha$ is calibrated to attain the prespecified
target-population outcome prevalence $p_0$.
Given $G$ and $X$, the observed outcome is determined by consistency.
We randomly sample $2{,}000$ cases and $2{,}000$ controls from the resulting
population pool and consider $p_0\in\{0.001,0.01,0.04,0.07,0.10\}$,
with 300 Monte Carlo replications per setting.
\par
\endgroup

We evaluate the two proposed estimators defined in Section~\ref{sec:estimation}. The first estimator, denoted by $\widehat\theta_E$, is based on the identification result in Theorem~\ref{thm:exact-identification} and requires the target-population disease prevalence $p_0=\mathbb{P}(Y=1)$ as an input. The second estimator, denoted by $\widehat\theta_A$, is based on the prevalence-free functional in Corollary~\ref{cor:rare-approximation}, which is motivated by the rare-outcome regime $p_0\approx0$ and does not require the value of $p_0$. 
Let $\theta_E$ and $\theta_A$ denote the true values targeted by $\widehat{\theta}_E$ and $\widehat{\theta}_A$, respectively. Their difference, $\theta_A-\theta_E$, represents a deterministic approximation error rather than sampling error. Both $\theta_E$ and $\theta_A$ are computed by Monte Carlo integration under the data-generating mechanism.
The nuisance functions are estimated using a convex-stacking ensemble with the learner library $\texttt{SL.glm}$, $\texttt{SL.glmnet.fast}$, $\texttt{SL.gam.fast}$, and $\texttt{SL.ranger.fast}$. Within each outer-training fold, three-fold cross-validation is used to select the convex-stacking weights.

%Full details on the simulation design and replication code are included
%in the Supplementary Material. Additional simulation exercises
%altering the overall sample size, specification of nuisance learners, and the sampling setting considered in Section 3.4 can be
%found in the Supplementary Materials

% We estimate the nuisance functions using five outcome-stratified outer folds. Within each outer-training sample, three-fold cross-validation selects convex-stacking weights for a library comprising main-effects logistic regression, elastic-net logistic regression with nonlinear and interaction terms, logistic GAM, and probability random forest. The Exact and Approx estimators use the same outer folds and out-of-fold nuisance predictions. We compute \(\theta_E\) and \(\theta_A\) from an independent reference population of \(2{,}000{,}000\) individuals and check both values by deterministic Gaussian integration. Supplementary Section S2 provides the complete computational audit.

\begin{figure}[!htbp]
  \centering
  \includegraphics[width=0.9\textwidth]{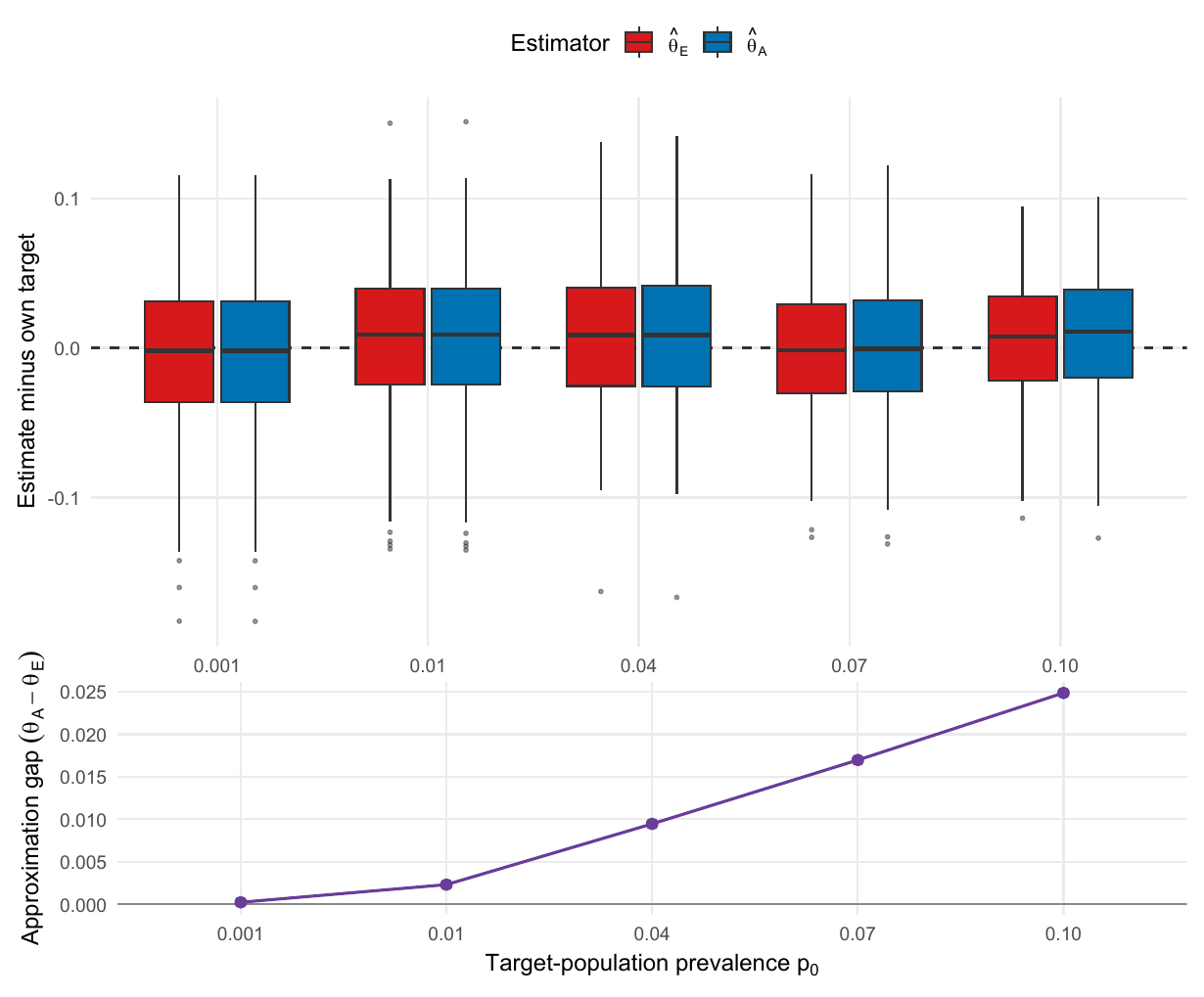}
  \caption{Boxplots of the $\PN$ estimates \(\widehat\theta_E-\theta_E\), \(\widehat\theta_A-\theta_A\) (upper panel) and the approximation gap $\theta_A-\theta_E$ (lower panel) with various disease prevalences $p_0$. }
  \label{fig:centered-errors-gap}
\end{figure}

%The upper panel shows \(\widehat\theta_E-\theta_E\) in red and \(\widehat\theta_A-\theta_A\) in blue. The lower panel shows the approximation gap \(\theta_A-\theta_E\).

\begin{table}[!htbp]
\centering
\begin{threeparttable}
\caption{Simulation results of two proposed estimators under various disease prevalences.}
\label{tab:estimand-aligned}
\scriptsize
\setlength{\tabcolsep}{12pt}
\begin{tabular}{llrrrrrr}
\toprule
$p_0$ & Estimator & Mean & Bias & SSD & ESE & RMSE & CP\\
\midrule
0.001 & \(\widehat\theta_E\) & 0.288 & -0.004 & 0.049 & 0.052 & 0.049 & 0.957\\
      & \(\widehat\theta_A\) & 0.288 & -0.004 & 0.049 & 0.052 & 0.049 & 0.957\\
0.010 & \(\widehat\theta_E\) & 0.298 &  0.006 & 0.049 & 0.051 & 0.049 & 0.957\\
      & \(\widehat\theta_A\) & 0.300 &  0.006 & 0.049 & 0.051 & 0.049 & 0.957\\
0.040 & \(\widehat\theta_E\) & 0.301 &  0.009 & 0.045 & 0.049 & 0.046 & 0.950\\
      & \(\widehat\theta_A\) & 0.311 &  0.010 & 0.046 & 0.050 & 0.047 & 0.943\\
0.070 & \(\widehat\theta_E\) & 0.291 & -0.001 & 0.042 & 0.048 & 0.042 & 0.970\\
      & \(\widehat\theta_A\) & 0.309 &  0.000 & 0.044 & 0.051 & 0.044 & 0.967\\
0.100 & \(\widehat\theta_E\) & 0.299 &  0.007 & 0.040 & 0.046 & 0.041 & 0.970\\
      & \(\widehat\theta_A\) & 0.326 &  0.009 & 0.043 & 0.049 & 0.044 & 0.970\\
\bottomrule
\end{tabular}
\begin{tablenotes}[flushleft]
\footnotesize
\item \textit{Abbreviations:} \(p_0\), the disease prevalence; Mean, the sample mean of the Monte–Carlo point estimates for 300 replications; Bias, the estimation bias relative to the corresponding true value; SSD, the empirical standard deviation of the point estimates; ESE, the averages of estimated asymptotic standard error based on Theorem~\ref{thm:asymptotic}; RMSE, root mean squared error; CP, the empirical coverage probability of 95\% Wald confidence intervals constructed using the EIF standard errors. Bias, RMSE, and CP are evaluated relative to \(\theta_E\) for \(\widehat\theta_E\) and \(\theta_A\) for \(\widehat\theta_A\).
\end{tablenotes}
\end{threeparttable}
\end{table}

%\begin{table}[!htbp]
%\centering
%\begin{threeparttable}
%\caption{Approximation and estimation components for \(\widehat\theta_A\).}
%\label{tab:gap-decomposition}
%\small
%\begin{tabular}{rrrrrr}
%\toprule
%\(p_0\) & \(\theta_E\) & \(\theta_A\) & \(\theta_A-\theta_E\) &
%Estimation bias & Total difference\\
%\midrule
%0.001 & 0.29182 & 0.29205 & 0.00023 & -0.00359 & -0.00336\\
%0.01  & 0.29182 & 0.29413 & 0.00231 &  0.00591 &  0.00822\\
%0.04  & 0.29202 & 0.30147 & 0.00945 &  0.00971 &  0.01916\\
%0.07  & 0.29203 & 0.30899 & 0.01696 &  0.00033 &  0.01729\\
%0.10  & 0.29198 & 0.31684 & 0.02486 &  0.00944 &  0.03429\\
%\bottomrule
%\end{tabular}
%\begin{tablenotes}[flushleft]
%\footnotesize
%\item Estimation bias is \(\E(\widehat\theta_A)-\theta_A\). Total difference is \(\E(\widehat\theta_A)-\theta_E\), exactly the sum of estimation bias and the approximation gap up to numerical precision.
%\end{tablenotes}
%\end{threeparttable}
%\end{table}
%Various disease prevalences \(p_0 = 0.10, 0.05, 0.01\) are considered.  Table~\ref{tab:estimand-aligned} reports the mean, bias, SSD, ESE, RMSE, and CP.and their SSDs generally decrease as \(p_0\) increases 
Figure~\ref{fig:centered-errors-gap} depicts the boxplots of \(\widehat\theta_E-\theta_E\) and \(\widehat\theta_A-\theta_A\) in the upper panel and the deterministic gap \(\theta_A-\theta_E\) in the lower panel.
Table~\ref{tab:estimand-aligned} summarize estimation results based on 300 Monte Carlo simulations. 
The Approx target $\theta_A$ is close to the Exact target $\theta_E$ at a relatively low prevalence $p_0 \le 0.10$. Specifically, at $p_0=0.10$, the approximation gap is $0.025$, corresponding to $8.5\%$ of $\theta_E$.
Relative to their target estimands, both estimators have small biases. The estimated asymptotic standard error ESE is slightly larger than the empirical standard deviation SSD across all settings, so the finite-sample intervals are mildly conservative, with CP ranging from 0.943 to 0.970.
%The values of \(\theta_E\) are approximately 0.292 across all settings, whereas \(\theta_A\) increases from 0.292 to 0.317.
%Table~\ref{tab:gap-decomposition} shows that the approximation gap increases from 0.00023 at \(p_0=0.001\) to 0.02486 at \(p_0=0.10\). At \(p_0=0.10\), \(\E(\widehat\theta_A)-\theta_E=0.03429\), comprising an estimation bias of 0.00944 relative to \(\theta_A\) and a deterministic approximation gap of 0.02486. 
%\FloatBarrier
\section{Application to the Global Enteric Multicenter Study}
\label{sec:application}

We demonstrate the practical utility of our methods in a real case--control study. The data are extracted from the Global Enteric Multicenter Study (GEMS), which is a comprehensive multi-center case--control study designed to systematically quantify the burden and discern the microbiologic etiology of diarrheal disease afflicting children aged 0--59 months in developing regions spanning Africa and Asia \citep{kotloff2012protocol,kotloff2013}. We focus on a total of 10,266 subjects across four African countries (The Gambia, Kenya, Mali, and Mozambique), including 4,266 qualifying moderate-to-severe diarrhea (MSD) cases and 6,000 controls. The outcome $Y$ of our interest is MSD (1 = MSD; 0 = non-MSD). We consider two microbial infections as exposures of interest, namely Cryptosporidium infection and Rotavirus infection, each coded as a binary exposure variable $X$ ($X=1$ for infected and $X=0$ for non-infected). Cryptosporidium was detected in 498 cases and 352 controls, and Rotavirus was detected in 752 cases and 287 controls. Several covariates are included to control for potential confounding, including sex, age, caregiver's education level, family wealth, body mass index (BMI), country, handwashing practices and water source, and contact with animals in residence. Different pathogen infections may confound one another. However, because the temporal ordering of these infections cannot be determined from the available data, directly adjusting for other infections may be inappropriate. Therefore, we do not include the other focal exposure or other recorded pathogens in the main adjustment set. In the Supplement material, we conduct sensitivity analyses to compare results with and without these additional infection variables; the substantive conclusions are unchanged. In this application, we estimate the probability of necessity (\PN) to quantify the probability that Cryptosporidium or Rotavirus infection was a causal contributor to moderate-to-severe diarrhea in African children.

Previous analyses based on the GEMS dataset have reported that both Cryptosporidium infection and Rotavirus infection are positively associated with diarrhea \citep{kotloff2013}. Existing clinical and epidemiologic evidence consistently suggests that these infections increase susceptibility to diarrheal disease rather than exert a protective effect. Therefore, the Monotonicity Assumption~\ref{ass:monotonicity} is a reasonable working assumption in this application. Moreover, even if monotonicity does not hold exactly, the prevalence-assisted functional $\theta_E$ provides a valid lower bound for $\PN$. Based on the literature review and expert experience, we specify that the prevalence of MSD is $p_0=0.08$.

First, we estimate \PN{} using two estimators: the estimator $\widehat\theta_A$ based on the rare-outcome approximation in Corollary~\ref{cor:rare-approximation}, and the estimator $\widehat\theta_E$ based on the exact identification result in Theorem~\ref{thm:exact-identification} with $p_0=0.08$. The corresponding results are presented in Table~\ref{tab:gems-main}. Two findings emerge from Table~\ref{tab:gems-main}. (1) For both pathogens, the estimate based on the rare-outcome approximation, $\widehat\theta_A$, is  close to the exact estimate, $\widehat\theta_E$. However, the confidence interval for $\widehat\theta_A$ is wider than that for $\widehat\theta_E$. The two intervals quantify sampling uncertainty for different functionals; their widths do not account for the approximation gap.  (2) For an infected child who developed MSD, the estimated probability that MSD would not have occurred in the absence of Cryptosporidium infection is 40.9\%, whereas the corresponding probability in the absence of Rotavirus infection is 73.9\%. These results suggest that, among infected children who developed MSD, Rotavirus is more likely than Cryptosporidium to have been a causal contributor to the disease.

\begin{table}[t]
\centering
\caption{Causal Attribution (PN) for Moderate-to-Severe diarrhea}
\label{tab:gems-main}
\begin{threeparttable}
\scriptsize
\setlength{\tabcolsep}{3pt}
\begin{tabular*}{0.88\textwidth}{@{\extracolsep{\fill}}lcc}
\toprule
Pathogen & $\widehat\theta_E$ (95\% CI) & $\widehat\theta_A$ (95\% CI) \\
\midrule
Cryptosporidium & 0.409 (0.331, 0.488) & 0.462 (0.369, 0.556) \\
Rotavirus & 0.739 (0.707, 0.771) & 0.821 (0.785, 0.857) \\
\bottomrule
\end{tabular*}
\end{threeparttable}
\end{table}

\begin{figure}[t]
\centering
\includegraphics[width=\textwidth]{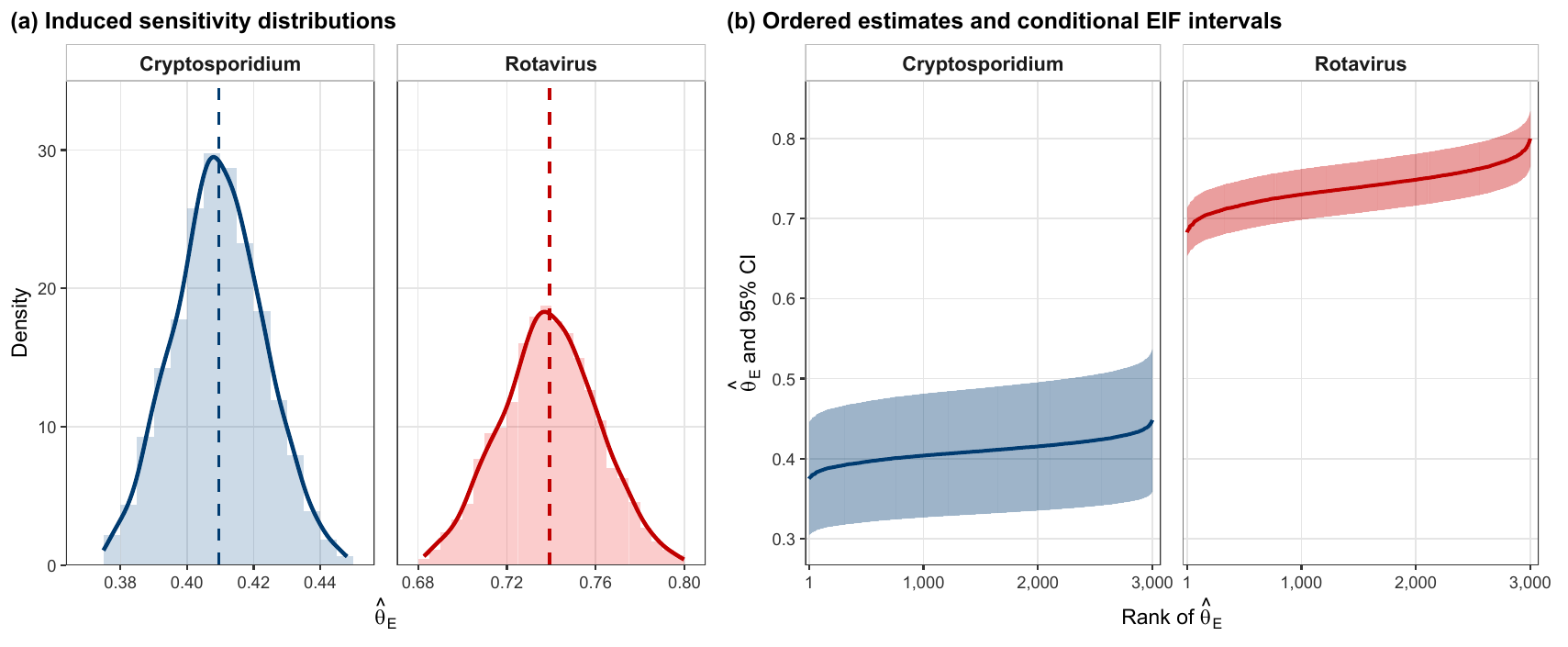}
\caption{Sensitivity analysis of pathogen-specific PN estimates $\widehat\theta_E$ under prior uncertainty in $p_0$. Panel (a) displays the empirical distribution of exact PN values obtained from repeated prior draws of $p_0$. Panel (b) displays the corresponding ordered exact PN estimates and their confidence intervals.}
\label{fig:gems-p0-probabilistic}
\end{figure}

Then, we conduct sensitivity analyses to evaluate the influence of the value of $p_0$. We use a prior-based probabilistic approach to quantify uncertainty. We specify a normal prior with mean $0.08$ and standard deviation $0.023$. We generate $p_0$ from the prior distribution and calculate $\mathrm{PN}$ estimates $\widehat\theta_E$. This experiment is repeated 3000 times.  The corresponding estimates are summarized in Figure~\ref{fig:gems-p0-probabilistic}: panel (a) presents histograms and density curves, and panel (b) presents the ordered estimates and their pointwise 95\% confidence intervals. All point estimates and confidence intervals are positive, supporting the robust positive effect of Rotavirus and Cryptosporidium infection on diarrhea risk. The estimated effects differ substantially between the two infections. Compared with Cryptosporidium infection, Rotavirus infection yields larger estimates and narrower confidence intervals.
% These distributions are sensitivity distributions induced by the stipulated distribution of $p_0$, not posterior distributions for \PN.

%\FloatBarrier

\section{Discussion}
\label{sec:discussion}

We develop a nonparametric framework for identification and efficient estimation of \PN{} under case--control sampling with measured confounding. The framework provides two approaches: an exact functional $\theta_E$ that identifies \PN{} under the stated causal assumptions and monotonicity using external population prevalence, and a simpler, prevalence-free approximation $\theta_A$ for rare outcomes. We characterize the approximation error and derive efficient influence functions for both targets, leading to cross-fitted estimators that accommodate flexible nuisance learning. Numerical studies illustrate how the proposed methods can address etiologic questions using retrospective data.
%The retrospective framework may also support identification and efficient estimation of other probabilities of causation.
 Without monotonicity, $\theta_E$ remains a valid lower bound for \PN{}. A further direction is to derive sharp causal bounds under case--control sampling and develop inference for their nonsmooth endpoints, extending recent work in randomized experiments \citep{zhang2026}. 
\bibliographystyle{plainnat}
\bibliography{references}
\end{document}